\documentclass[12pt]{article}
\usepackage{titling}
\usepackage[utf8]{inputenc}
\usepackage{amsmath}
\usepackage{stmaryrd} 
\usepackage{mathtools} 
\usepackage{listings} 
\usepackage{tikz}
\usepackage{tkz-euclide}
 \usepackage{cancel}
\usepackage[french]{babel}
\usepackage[T1]{fontenc}  
\usepackage{csquotes}
\usepackage[backend=biber,maxbibnames = 4,sorting=none]{biblatex} 
\usepackage{url}
\title{Infomathic}

\usetikzlibrary{calc,arrows.meta}

\usepackage{pgfplots}

\usepackage{mathpazo}
\usepackage{amsmath}

\usepackage{pifont}

\author{Karim \textsc{Zayana}\up{1,2}, Régis \textsc{Quéruel}\up{1}, Pierre \textsc{Michalak}\up{1}}

\date{\small\up{1} Ministère de l'\'Education nationale, Paris\\
\up{2} LTCI, Télécom Paris, Institut Polytechnique de Paris}
\begin{document}


\maketitle
Depuis qu'il existe, l'outil informatique a souvent épaulé les mathématiciens, qu'il s'agisse d'implémenter une méthode d'approximation (calcul numérique d'une racine, d'une intégrale,\ldots) ou de simuler un phénomène (de nature géométrique, probabiliste,\ldots) pour vérifier ou établir une conjecture. Les programmes scolaires placent d'ailleurs très tôt (dès le collège) ces aspects essentiels du dialogue entre \textbf{info}r\textit{math}iciens parmi leurs intentions majeures  \cite{progCollege,progSeconde,zayana2018informathique}. 

\bigskip
Mais, et c'est un autre point sur lequel nous concentrerons ici notre attention, l'informatique aura également servi la cause des mathématiques en inspirant certains raisonnements ou en prenant à sa charge les pans entiers d'une démonstration. Nous allons illustrer ce fructueux partenariat par deux exemples  accessibles dès la classe de première. 

\section{Grille de nombres à classer}
On considère un tableau numérique au format $3\times 5$ (soit $3$ lignes et $5$ colonnes), par exemple la matrice $3\times 5$
\[
T = \begin{bmatrix}
1&8&3&4&8\\
0&9&2&7&14\\
20&3&6&7&7
\end{bmatrix}.
\]
Ligne par ligne, on dispose les nombres dans l'ordre croissant (de la gauche vers la droite), ce qui génère 
\[
T' = \begin{bmatrix}
1&3&4&8&8\\
0&2&7&9&14\\
3&6&7&7&20
\end{bmatrix}.
\]
On procède de même avec les colonnes (de haut en bas), ce qui produit
\[
T'' = \begin{bmatrix}
0&2&4&7&8\\
1&3&7&8&14\\
3&6&7&9&20
\end{bmatrix}.
\]

On observe que les lignes ont changé, certes, mais qu'elles demeurent classées du plus petit élément au plus grand. On en voudrait la preuve\ldots laquelle n'est pas si simple, du moins pas autant que le résultat énoncé ! Pour ce faire, nous définirons $T = [t_{i,j}]_{1\leq i \leq n,1\leq j \leq p}$ au format $n\times p$ et supposerons que ses lignes sont préalablement ordonnées, à savoir que, d'ores et déjà, $T=T'$. Nous distinguerons d'abord le cas $n=2$.

\bigskip
\textbf{Cas d'une matrice à deux lignes ($n=2$)}. Voyons pour commencer le cas d'un tableau biligne, soit 
\[
T=T'= \begin{bmatrix}
t_{1,1}&\leq&t_{1,2}&\leq&t_{1,3}&\ldots&\leq&t_{1,p}\\
t_{2,1}&\leq&t_{2,2}&\leq&t_{2,3}&\ldots&\leq&t_{2,p}
\end{bmatrix}.
\]
où les symboles d'inégalités placés entre les coefficients ne sont apparents que pour rappeler l'ordre qui les lie. Compte-tenu des contraintes imposées sur les colonnes à bâtir,
\[
T'' = \begin{bmatrix}
\min(t_{1,1},t_{2,1})&\min(t_{1,2},t_{2,2})&\ldots&\min(t_{1,p},t_{2,p})\\
\max(t_{1,1},t_{2,1})&\max(t_{1,2},t_{2,2})&\ldots&\max(t_{1,p},t_{2,p})
\end{bmatrix}.
\]

La retouche pratiquée sur les colonnes ne perturbe pas l'ordonnancement des lignes. En effet,
\begin{itemize}
\item $\min(t_{1,1},t_{2,1})\leq t_{1,1}$, lui-même inférieur à $t_{1,2}$;
\item $\min(t_{1,1},t_{2,1})\leq t_{2,1}$, lui-même inférieur à $t_{2,2}$;
\item donc\ldots $\min(t_{1,1},t_{2,1})\leq \min(t_{1,2},t_{2,2})$;
\item \ldots Ainsi s'enchaîne la première ligne.
\end{itemize}
Symétriquement,
\begin{itemize}
\item $\max(t_{1,2},t_{2,2})\geq t_{1,2}$, lui-même supérieur à $t_{1,1}$;
\item $\max(t_{1,2},t_{2,2})\geq t_{2,2}$, lui-même supérieur à $t_{2,1}$;
\item donc\ldots $\max(t_{1,1},t_{2,1})\leq \max(t_{1,2},t_{2,2})$;
\item \ldots Ainsi s'enchaîne la deuxième ligne.
\end{itemize}

\bigskip
\textbf{Cas général}. Reprenons maintenant le cas de notre matrice $T=T'$ de taille $n\times p$. À la manière d'un tri bulle \cite{cormen1994algo}, dont on s'apprête à mimer le mécanisme en l'étendant d'une dimension, échangeons les coefficients qui le doivent entre la ligne $1$ et la ligne $2$, la (nouvelle) ligne $2$ et la ligne $3$,\ldots, la (nouvelle) ligne $n-1$ et la ligne $n$. À l'issue de ce premier passage,
\begin{itemize}
\item toutes les lignes restent bien ordonnées : cela relève du cas précédent ($n=2$);
\item la dernière ligne a pris son aspect définitif -- elles ne comporte que les <<~majors~>> de chaque colonne, mais pas (encore) les précédentes. Il faudra donc reparcourir la matrice par couples de lignes consécutives, en s'arrêtant cette fois à l'avant-dernière -- pour que celle-ci qui se fige, à l'antépénultième, etc.
\end{itemize}
Le processus aura fait son \oe{}uvre après le $n-1$ ième passage. Inutile de l'encoder : ici tout est \og débranché \fg.

\section{Suite de Porges}
Authentique joyau, ce problème est un pur jeu de l'esprit \cite{porges1945set,olympiades2017sujet}. Prenons un entier naturel $n$. Associons-lui la somme des carrés de ses chiffres en base $10$. Par exemple, $n=0$ devient $0^2=0$, $n=12$ devient $2^2+1^2=5$, $n=308$ devient $8^2+0^2+3^2 = 73$. Formellement, nous avons défini l'application $f$ qui transforme le nombre générique $n$ composé de $p$ chiffres,
$$
n=a_{p-1} 10^{p-1} + a_{p-2} 10^{p-2}+\cdots+a_1 10 + a_0
$$
où $p\geq0$ et $a_i\in\llbracket0,9\rrbracket$ où $0\leq i \leq p-1$, en le nombre
$$
f(n) = a_{p-1}^2+\cdots +a_1^2+a_0^2.
$$
Cette définition est univoque : les zéros bordant éventuellement et gratuitement sur la gauche l'écriture décimale de $n$ n'affecteraient pas la somme.

\bigskip
On peut d'ores et déjà dégager quelques propriétés de $f$. Elle est surjective car tout entier $p$ est l'image de l'entier $11\cdots1$ qui s'écrit avec $p$ chiffres $1$. Elle n'est pas injective car plusieurs entiers peuvent avoir la même image, par exemple $1$ et $10$. Surtout, et ce sera l'objet de notre étude, on constate à l'usage qu'appliquer sans répit $f$ à un naturel $n$ donné semble toujours aboutir, au choix, 
\begin{itemize}
\item sur le naturel $0$. Auquel cas, on en était parti et on n'en bouge plus.
\item sur le naturel $1$. Auquel cas on n'en bouge plus. On qualifie cet événement d'\og heureux \fg.
\item sur le naturel $4$. Auquel cas s'amorce un cycle de longueur $8$, où défilent successivement $f(4) = 16$, $f(16) = 37$, $f(37)= 58$, $f(58)=89$, $f(89)=145$, $f(145)=42$, $f(42)=20$ avant de reboucler sur $4=f(20)$.
\end{itemize}

Cette conjecture n'est pas sans rappeler celle de Syracuse \cite{conway2013syracuse}. Mais, contrairement à cette dernière, elle se démontre ! La preuve se structure en trois étapes. 

\bigskip
\textbf{Nombre $n$ à un ou deux chiffres}. C'est ici que l'informatique nous est d'un grand secours : on y contrôle à la main la propriété pour chaque entier $n$ de l'intervalle $\llbracket0,99\rrbracket$, par la \og force brute \fg\; donc. L'ordinateur nous en soulage des laborieux calculs. On code d'abord la fonction $f$ recevant l'argument $n$ et retournant $f(n)$, par exemple en Python :
\begin{lstlisting}
def f(n):
    return sum( int(chiffre)**2 for chiffre in str(n) )
\end{lstlisting}

Puis on teste la propriété sur tous les entiers de $0$ à $99$, ce qui la validera sur cette plage. On attend que la boucle termine, ce qu'elle finit par faire. Sur une machine X250 de marque Lenovo, le temps d'exécution avoisine le millième de seconde.
\begin{lstlisting}
attracteurs = [0,1, 4, 16, 37, 58, 89, 145, 42, 20]
for n in range(0, 99):
    m = n
    while m not in attracteurs:
         m = f(m)
print("verification faite")
\end{lstlisting}

\textbf{Nombre $n$ à trois chiffres}.
On se ramène progressivement au cas précédent par un élégant subterfuge. Soit $n$ un entier compris entre $100$ et $999$. Il s'écrit $n=100a+10b+c$ avec $a$, $b$, $c$ dans $\llbracket0,9\rrbracket$, $a$ non nul. Remarquons que 
$$
n-f(n) = a(100-a)+b(10-b)+c-c^2.
$$
La quantité $a(100-a)$ correspond aux ordonnées des points d'une branche de parabole dans sa portion croissante puisque $\llbracket1,9\rrbracket \subset ]-\infty,50]$. Si bien que pour tout $a$ de $\llbracket1,9\rrbracket$, $a(100-a) \geq 1\times(100-1) =99$. De plus $b(10-b) \geq 0$ et, sans rechercher la précision, $99+c-c^2 \geq 99+0-81=18$. A fortiori, $n-f(n)\geq 1$. Dès lors 
$$
f(n) \leq n-1.
$$  
En itérant suffisamment, on descend sous la barre des $100$.

\bigskip
\textbf{Nombres à quatre chiffres et au-delà}
On se ramène, là encore patiemment, aux cas précédents. En effet, si
\[
n=a_{p-1} 10^{p-1} + a_{p-2} 10^{p-2}+\cdots+a_1 10 + a_0
\]
avec $a_{p-1} \neq 0$, nous avons aisément
\[
f(n)= a_{p-1}^2+\cdots +a_1^2+a_0^2 \leq 81p.
\]
Or $81p$ est lui-même inférieur (strictement) à $10^{p-1}$, par récurrence\footnote{Observer que $81(p+1)=81p + 81$, que $81\leq 10^{p-1}$ puis pour l'hérédité que $81p+81$ est inférieur à $10^{p-1} + 10^{p-1}$ lui-même strictement inférieur à $10^p$.} sur $p$. On en retient que $f(n)$, qui occupait $p$ chiffres, ne se déploie plus que sur $p-1$ chiffres (au plus). En persévérant, on retrouve les situations à $3$, $2$ ou $1$ chiffre(s), déjà examinées plus haut. 

\bigskip
Arthur \textsc{Porges} (1915 -- 2006), l’auteur de l’article  <<~A Set Of Eight Numbers~>> dont nous nous sommes inspirés, fut aussi essayiste et poète. Né à Chicago le 20 août 1915, il obtient d’abord une maîtrise en mathématiques à l'Illinois Institute of Technology. Après avoir servi dans l'armée américaine en tant qu'instructeur de mathématiques pendant la Seconde Guerre mondiale (époque où il publie dans \textit{The American Mathematical Monthly}), il déménage en Californie et enseigne dans différents établissements, dont le Los Angeles City College. Fervent lecteur de fictions, Porges commence à rédiger ses propres nouvelles. En 1950, il signe la première, <<~Modeled in Clay~>>, écrite dans le genre de la <<~fantasy~>>, à la croisée du merveilleux et du fantastique. Fort de ce succès, il persévère, toujours en marge de son travail quotidien de professeur. En 1957, il se retire de l'enseignement et devient rapidement un écrivain accompli et reconnu, en particulier dans l’univers de la science-fiction.
\begin{figure}[htbp]
\centering
\includegraphics[width =.7\linewidth]{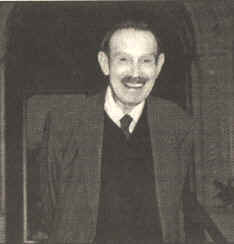}
\caption{Arthur Porges (1915 -- 2006)}
\label{uneroue}
\end{figure}
\section*{Remerciements}
Les auteurs remercient le comité éditorial de la revue Quadrature ainsi que Christine \textsc{Weill}, inspectrice d'académie -- inspectrice pédagogique régionale de mathématiques dans l'académie de Versailles, pour leurs relectures et échanges autour de ce texte, sans oublier George \textsc{Sicherman}, Richard \textsc{Simms} et Antoine \textsc{Porgès} pour leur aide à reconstituer la biographie d'Arthur \textsc{Porges}. 

\printbibliography 
\end{document}